**Title : Electron-Scale Dynamics of the Diffusion Region during Symmetric Magnetic Reconnection in Space**

R. B. Torbert[1,2], J. L. Burch[2], T. D. Phan[3], M. Hesse[4,2], M. R. Argall[1], J.Shuster[5], R. E. Ergun[6], L.Alm[7], R. Nakamura[8], K. Genestreti[8], D. J. Gershman[5], W.R. Paterson[5], D. L. Turner[9], I. Cohen[10], B. L. Giles[5], C. J. Pollock[5], S.Wang[11], L.-J. Chen[11], Julia Stawarz[12], J. P. Eastwood[12], K.-J.Hwang[2], C. Farrugia[1], I. Dors[1], H. Vaith[1], C.Mouikis[1], A. Ardakani[1], B. H. Mauk[10], S. A. Fuselier[2,13], C. T. Russell[14], R. J. Strangeway[14], T.E. Moore[5], J. F. Drake[10], M. A. Shay[15], Yu. V. Khotyaintsev[7], P.-A. Lindqvist[16], W. Baumjohann[8], F. D. Wilder[6], N. Ahmadi[6], J. C. Dorelli[5], L. A. Avanov[5], M. Oka[3], D. N. Baker[6], J. F. Fennell[9], J. B. Blake[9], A. N. Jaynes[17], O. Le Contel[18], S. M. Petrinec[19], B. Lavraud[20], Y. Saito[21]

**Affiliations:**
[1] University of New Hampshire, Durham, New Hampshire, USA
[2] Southwest Research Institute, San Antonio TX, USA
[3] University of California, Berkeley, USA
[4] University of Bergen, Norway
[5] NASA, Goddard Space Flight Center, Greenbelt, Maryland, USA
[6] University of Colorado LASP, Boulder, Colorado, USA
[7] Swedish Institute of Space Physics, Uppsala, Sweden
[8] Space Research Institute, Austrian Academy of Sciences, Graz, Austria
[9] Aerospace Corporation, El Segundo, California, USA
[10] Johns Hopkins University Applied Physics Laboratory, Laurel, Maryland, USA
[11] University of Maryland, College Park, Maryland, USA
[12] Blackett Laboratory, Imperial College London, London, UK
[13] The University of Texas at San Antonio, San Antonio, Texas, USA
[14] University of California, Los Angeles, Los Angeles, California, USA
[15] University of Delaware, Newark, Delaware, USA
[16] Royal Institute of Technology, Stockholm, Sweden
[17] University of Iowa, Iowa City, Iowa USA
[18] Laboratoire de Physique des Plasmas, CNRS/Ecole Polytechnique/Sorbonne Université/Univ. Paris Sud/Observatoire de Paris, Paris, France.
[19] Lockheed Martin Advanced Technology Center, Palo Alto, California, USA
[20] Institut de Recherche en Astrophysique et Planétologie, CNRS, CNES, Université de Toulouse, France
[21] ISAS/JAXA, Japan

**One Sentence Summary:** NASA's Magnetospheric Multiscale mission detected fast magnetic reconnection and high-speed electron jets in the Earth's magnetotail.

**Abstract:** Magnetic reconnection is an energy conversion process important in many astrophysical

contexts including the Earth's magnetosphere, where the process can be investigated in-situ. Here

we present the first encounter of a reconnection site by NASA's Magnetospheric Multiscale (MMS)

spacecraft in the magnetotail, where reconnection involves symmetric inflow conditions. The unprecedented electron-scale plasma measurements revealed [1] super-Alfvénic electron jets reaching 20,000 km/s, [2] electron meandering motion and acceleration by the electric field, producing multiple crescent-shaped structures, [3] spatial dimensions of the electron diffusion region implying a reconnection rate of ~0.1-0.2. The well-structured multiple layers of electron populations indicate that, despite the presence of turbulence near the reconnection site, the key electron dynamics appears to be largely laminar.

**Main Text:** Investigation of magnetic reconnection requires ultra-high resolution plasma and field measurements to resolve electron-scale structures inside the electron diffusion region, or EDR (*1*). Although previous spacecraft missions have encountered the EDR (*2-4*), MMS is uniquely capable of electron distribution measurements at resolution 100 times better than before. MMS focuses on two important reconnection regions, the dayside magnetopause and the nightside magnetotail, with very different plasma parameter regimes. During its first phase (2015-2016), MMS investigated dayside magnetopause reconnection (*1*), where the inflow conditions are highly asymmetric (with different plasma and magnetic pressures in the two inflow regions), and magnetic energy conversion processes occur in two separated regions--the X-line, where the magnetic field reverses, and the electron flow stagnation point (*5-6*). In its second phase (2017), MMS explored the kinetic processes of reconnection in the Earth's magnetotail where the inflow conditions are nearly symmetric, the available magnetic energy per particle is more than an order of magnitude higher than at the dayside, and the X-line and stagnation point are coincident (*7*). Since the amount of magnetic energy per particle in the magnetotail is comparable to that of the solar corona, what is learned in the tail has broader implications. This report presents the first detailed investigation of electron physics in magnetotail reconnection, verifying some predictions, while discovering new, yet unexplained features.

On 11 July 2017, at ~22:34 UT, MMS encountered an EDR when it detected tailward-directed plasma flows followed by earthward-directed flows (Figures 1f,g), spanning a reversal of the north-south component of the magnetic field, $B_z$ (1d). The spacecraft were at a radial distance of 22 Earth radii in the magnetotail (see **S1**). Four-spacecraft timing of the flow and field reversals indicate that the structure moved away from Earth with velocity $V_x$~ -190 km/s. These are classical signatures of the tailward retreat of the reconnection X-line past the spacecraft (Figure 1j) (*e.g.,3,4,8-13*). In this case, except for a brief excursion to the edge of the inflow region, seen in a small perturbation in **B** beginning at 22:34:00UT (due to a flapping of the current sheet), the spacecraft stayed close to the neutral sheet, indicated by small values of $|Bx|$ ( ~0-2 nT), during the flow and field reversal. These observations are consistent with crossing both ion and electron diffusion regions, further supported by the profiles of the ion and electron flows: the out-of-plane electron velocity, $V_y$, peaked at 20,000 km/s, the order of the electron Alfvén speed. Starting from the X-line (at the **V** and **B** reversal location) and going left and right in Figure 1h, the electron outflow speed $|Vex|$ increased and greatly exceeded the ion speed. While the ion outflow speed $|Vi_{\perp}x|$ increased with increasing distance from the X-line, $|Ve_{\perp x}|$ reached a peak (~7,000 km/s) before slowing down and approaching the ion flow speed at ~22:33:50 before, and ~22:34:20 after, the X-line. Thus the ends of the ion diffusion region, where the ion and electron outflow velocities are expected to match, are likely observed near these times. The ends of the electron diffusion region, on the other hand, marked by the departure of $Ve_{\perp}$ from **E**x**B**$/B^2$, was confined to a much smaller interval around the X-line, where the electron density reached a symmetric minimum of 0.03 cm$^{-3}$ (electron inertial length, $d_e \sim 30$ km). To orient the data within the EDR to the nominal picture of reconnection, as in Figure 1j, a boundary normal (LMN) coordinate system was determined (see **S1**).

Figures 2a-j (and 3a-e) display MMS3 data in and around the EDR in these coordinates. Figures 2(k-n) show reduced electron distribution functions (DF's) during the strong reconnecting current ($J_M$) at times indicated by the vertical dashed lines. These times are before, and at, the peak of the quantity $\mathbf{J} \cdot \mathbf{E}'$ (where $\mathbf{E}'=\mathbf{E}+ \mathbf{V}_e\times\mathbf{B}$), which is the electromagnetic energy conversion rate in the plasma frame, one signature of the EDR (*14*). Although $\mathbf{J} \cdot \mathbf{E}'$ is mostly positive throughout the plotted interval, there are some regions with significant negative values, indicating that the electrons are transferring energy to the electromagnetic field, as seen also in simulations (*14-16*). Figure 3c shows that, at all spacecraft, the signs of $E_N$ and $B_L$ were anti-correlated, consistent with $E_N$ converging toward the neutral sheet ($B_L$=0) from both hemispheres, as expected for symmetric reconnection with a minimal guide field (*10,12,17,18*). MMS2 (and 4) remained below the neutral sheet ($B_L$<0 and $E_N$>0) in the vicinity of the EDR crossing, while MMS1 and 3, located at higher N (or ~$Z_{GSM}$), made excursions above, where $B_L$>0 and $E_N$<0. This $E_N$ field accelerates the neutral sheet electrons towards the inflow region where they are accelerated along meandering trajectories (*19*) by the reconnection field, $E_M$ ~1-2 mV/m (Figures 3c,e and see **S1**). The electrons eventually turned into the L, or exhaust, direction by $B_N$ as they exited the EDR, forming the electron jet seen in figures 2c and 3b on either side of the X-line.

The electron temperature profile in panel 2f shows strong anisotropy from 22:34:01.0 to 01.8 due to magnetic field-aligned electrons in the in-flow region (*2*). During the EDR crossing, there was only a small rise (few 100 eV) in parallel (or perpendicular) temperature, unlike the case of asymmetric reconnection (1), implying that a substantial fraction of the energy conversion went into the strong electron flows in the M and L directions.

The aspect ratio of the EDR is an important reconnection parameter that has never been determined experimentally but has been a focus of theoretical and simulation studies for many years (*2,17,20*). Four-spacecraft timing analysis of the $B_N$ reversal near 22:34:02.2 (see Figure 2a) indicates that the X-line structure was moving tailward ($V_{XL} \sim$ -170 km/s). The EDR length can be estimated by multiplying $V_{XL}$ by the 1/e width of $Ve_M$ (~3s, Figure 2c), or by the $|Ve_L|$ peak-to-peak time (~2s, Figure 2d), yielding a full length of 350-500 km (12-17 $d_e$). MMS also made a brief excursion into the EDR inflow region (beginning at ~22:34:01.0), indicated by the increase in $|B_L|$ and confirmed by the cooler electrons (Figure 2b). By 22:34:02.2, the change in $B_L$ and the timing analysis ($V_{XN} \sim$ -70 km/s) show the structure moving southward, giving MMS also a normal motion into the EDR, reaching the neutral sheet and the peak of the cross-tail current by 22:34:03.0. Using Ampere's Law, dividing the change in $B_L$ during this normal motion into the EDR (Figure 2a, ~ 22:34:02.0 to 03.0) by the average of $J_M$, yields a simple estimate of the normal half-width of 30 km,~1 $d_e$ (see also **S2**). Thus, the aspect ratio (i.e.,reconnection rate) is ~0.1 - 0.2, consistent with fast reconnection (*21*).

Multiple crescent and triangular-shaped features in the DF's (Figures 2(k-n) and lower panels of Figure 3) are the result of electron meandering motion in the electromagnetic field structure of the EDR. Figure 2l shows a DF taken at a location below (in N) the EDR, which features multiple crescents, seen as enhanced phase space density at increasing velocities, quite similar to those predicted earlier (*22-24*) and shown in Figure 2q from the simulation of Figure 2o (see **S3**). Contrary to magnetopause observations and models (*1,25*), we here find more than one crescent. Furthermore, the observations show that crescents at higher $V_{\perp 1}$ are broader in $V_{\perp 2}$ than models predict. Models do show that these crescents are generated by the interaction of bouncing electrons with both the normal ($E_N$) and the reconnection electric field ($E_M$). The fact that observations show multiple crescents indicates that the rather complex electron orbits are relatively unperturbed by high frequency fluctuations in the electromagnetic fields.

Figures 2(m,n) display a second DF, taken near the X point, which features a pronounced triangular shape in the plane containing B. Figure 2n shows two significant enhancements at lower $V_{\perp 1}$ (at +/-$V_{\perp 2}$, essentially $V_N$) indicating the inflowing populations from both above and below the X point. This distribution is similar to those predicted from simulation, figure 2s. In figure 2m, the triangular shape narrows in width as energy increases, again similar to simulation, figure 2r. Bouncing electrons account for this feature: for each bounce, electrons gain successively more energy from acceleration by the reconnection electric field. If electrons feature a finite $V_L$, they eventually interact with the magnetic field in the outflow, and are ejected from the immediate vicinity of the X point. The acceleration by the reconnection electric field and this ejection explain the triangular shape of the distribution: only electrons with very small $V_L$ remain near the X point long enough to execute multiple bounces and be accelerated to higher energies.

The electron DFs of Figure 3 (lower panels) show the evolution of the above features as MMS entered the EDR. From signatures of the inflow region (*2*), with DFs elongated along B, (MMS1 and 2, first column, at 22:34:02.514), the spacecraft, with MMS3 leading, penetrated farther into the current layer and saw accelerated and gyrating electrons growing in energy as time (and N position) increased, showing a perpendicular crescent with energy >1 keV ($2 \times 10^4$ km/s). By 22:34:02.724, all SC spacecraft were showing the perpendicular crescents, enhanced flow along the **ExB** direction, and also beaming features in the parallel directions. The parallel beams may be responsible for the high frequency electrostatic noise near the upper hybrid frequency (~1200 Hz), seen at this time in Figure 2i (*26*). When the spacecraft were fully within the reconnecting current layer (panel b), there were higher energy swirling features bending into both the $V_{\perp 1}$ (~M) and the $V_\parallel$ directions along with persistent counterstreaming, low energy($\sim$10,000 km/s)  field-aligned beams. By 22:34:02.787, MMS3, which was deepest in the EDR, saw very energetic electrons in $V_{\perp 1}$, and also in the $-V_\parallel$ direction: i.e., these accelerated electrons were rapidly leaving the EDR region. The evolution of many such features can be seen in movie **S1**.

We presented the first MMS observations of the magnetotail reconnection electron diffusion region which differs from that on the dayside as it involves symmetric inflow. These observations have yielded critical new insights into the nature of the EDR. MMS determined the aspect ratio of the diffusion region and hence a reconnection rate of 0.1-0.2, consistent with many simulations (*7,14,16,23*). MMS also revealed that the electron dynamics in the diffusion region matches well the predictions made by one class of theories and models – namely the ones that assume that effects of turbulence and associated fluctuations on the electron dynamics are small. Contrary to the magnetopause results (*1*), we find here that electrons can be accelerated up to three successive times by the reconnection electric field- possibly a consequence of better confinement in the symmetric magnetic structure. Taken together with MMS observations at the magnetopause, these new results provide a discriminator among competing theories of reconnection.  Some apparent differences between these observations and models remain to be understood, such as the energy width of the electron crescents.

**Data Availability**

The entire MMS data set is available on-line at https://lasp.colorado.edu/mms/sdc/public/links/, with modifications described in **S2**.

## Acknowledgements

The dedicated efforts of the entire MMS team are greatly appreciated. We are especially grateful to the leadership of the GSFC Project Manager, the late Craig Tooley, his Deputy, Brent Robertson, and the SwRI Payload Project Manager, Ron Black. This work was supported by NASA prime Contract No. NNG04EB99C at SwRI, and others cited in **S5**.

**Figures**

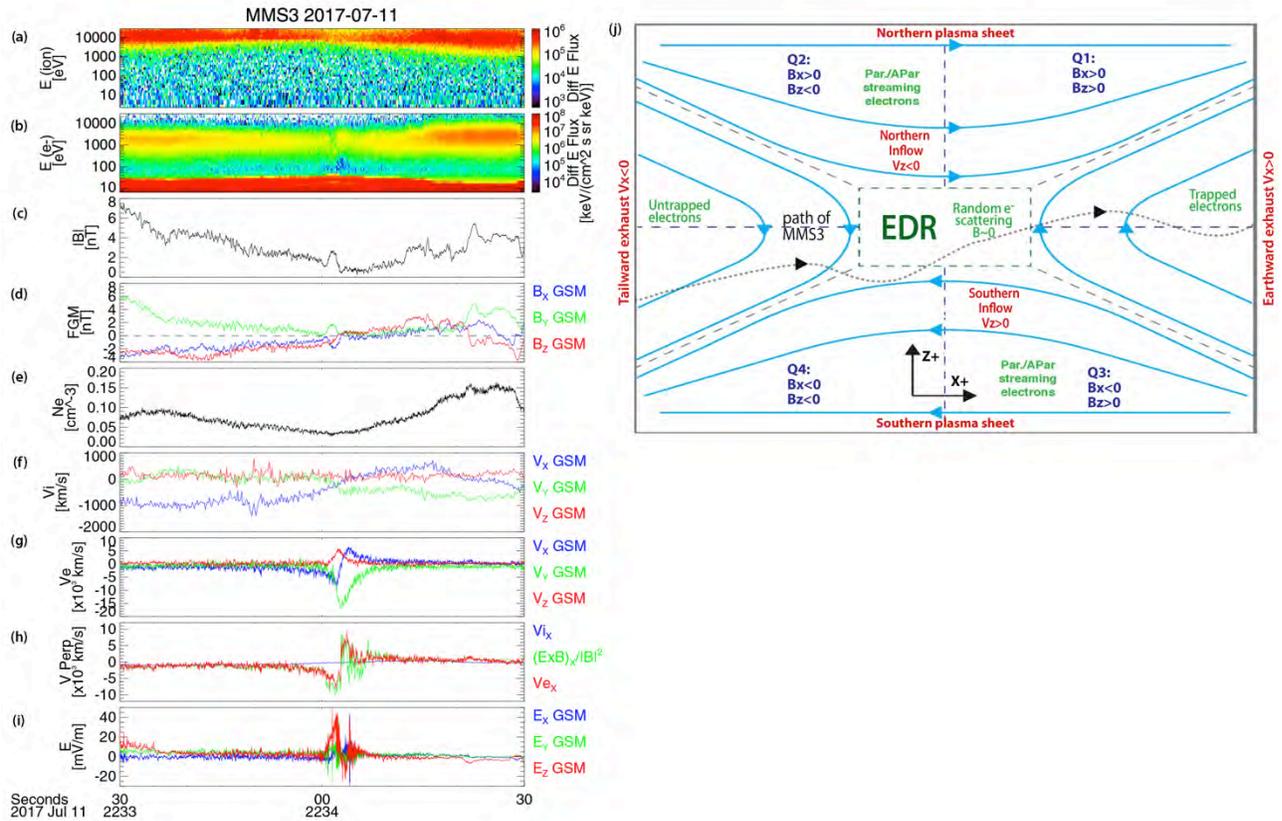

**Figure 1. MMS3 summary data** near the crossing of the EDR at 223403 on 2017 July 11. Panel data include: (a,b) energy-time spectrograms of (ion,electron) energy flux; (c) magnetic field magnitude, and (d) components in the GSM coordinate system; (e) electron density; (f) ion bulk velocity vector; (g) electron bulk velocity; (h) the x-component of perpendicular ion and electron flow, and of **E**x**B**/B$^2$ ; (i) electric field. The diagram to the right (j) is an illustration of a typical symmetric EDR, and the expected properties in various quadrants (Q), together with the inferred relative path of the MMS satellites as the X-line retreated tailward.

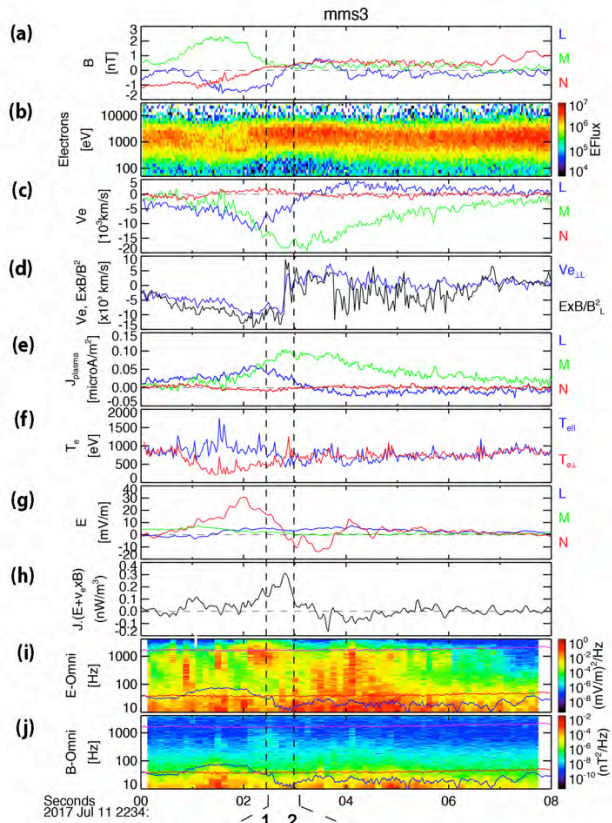

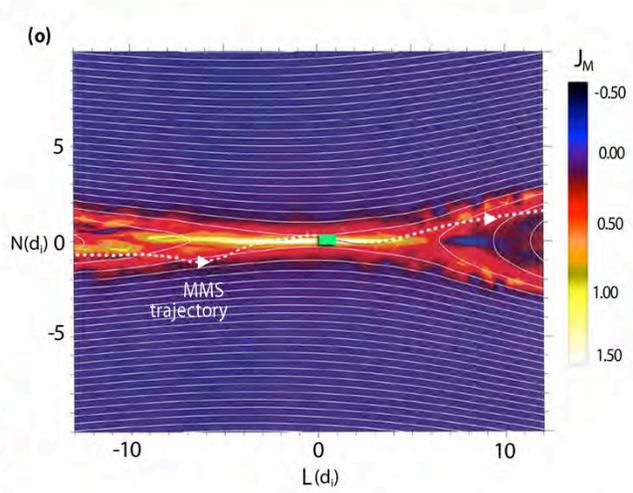

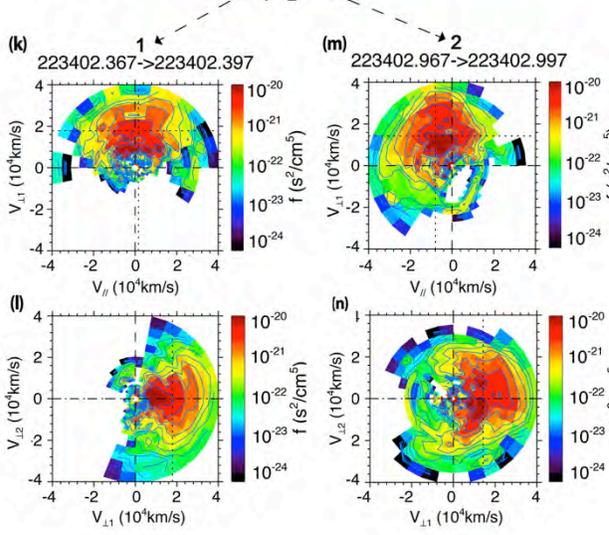

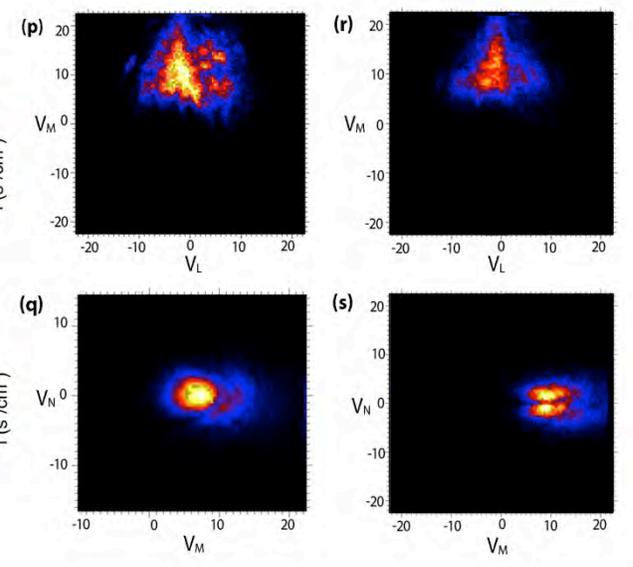

**Figure 2. MMS3 plasma and field data on 2017 July 11.** (a) Magnetic field components in LMN coordinate system; (b) electron omni-directional spectrogram, with minimum energy set at 50 eV to avoid the lower energy spacecraft photoelectrons seen in Figure 1b; (c) electron bulk velocity; (d) L components of $\mathbf{V}e_\perp$ and $\mathbf{E}x\mathbf{B}/B^2$ ; (e) Current from plasma measurements; (f) $Te_{||}$ and $Te_\perp$; (g) Electric field; (h) $\mathbf{J} \cdot \mathbf{E'}$ ; (i,j) (Electric, magnetic) omni-directional frequency spectrograms; (k,l,m,n) Electron velocity distribution functions at times indicated, $V_{\perp 1}$ being $(\mathbf{b} \times \mathbf{v}) \times \mathbf{b}$ , where $\mathbf{b}$ and $\mathbf{v}$ are unit vectors of $\mathbf{B}$ and $\mathbf{V_e}$; $V_{\perp 2} = \mathbf{v} \times \mathbf{b}$ ; and $V_{||}$.  $V_{\perp 1}$ is essentially the $\mathbf{E} \times \mathbf{B}$ direction and the bulk flow component in that direction is indicated by the faint dashed vertical lines; (o) Magnetic configuration of a computer simulation (**S3**), with color-coded reconnection current ($J_M$) ;(p,q,r,s) Reduced distribution $f_e$ near the green box in (o) from that simulation, with velocity axes corresponding to data panels k-n.

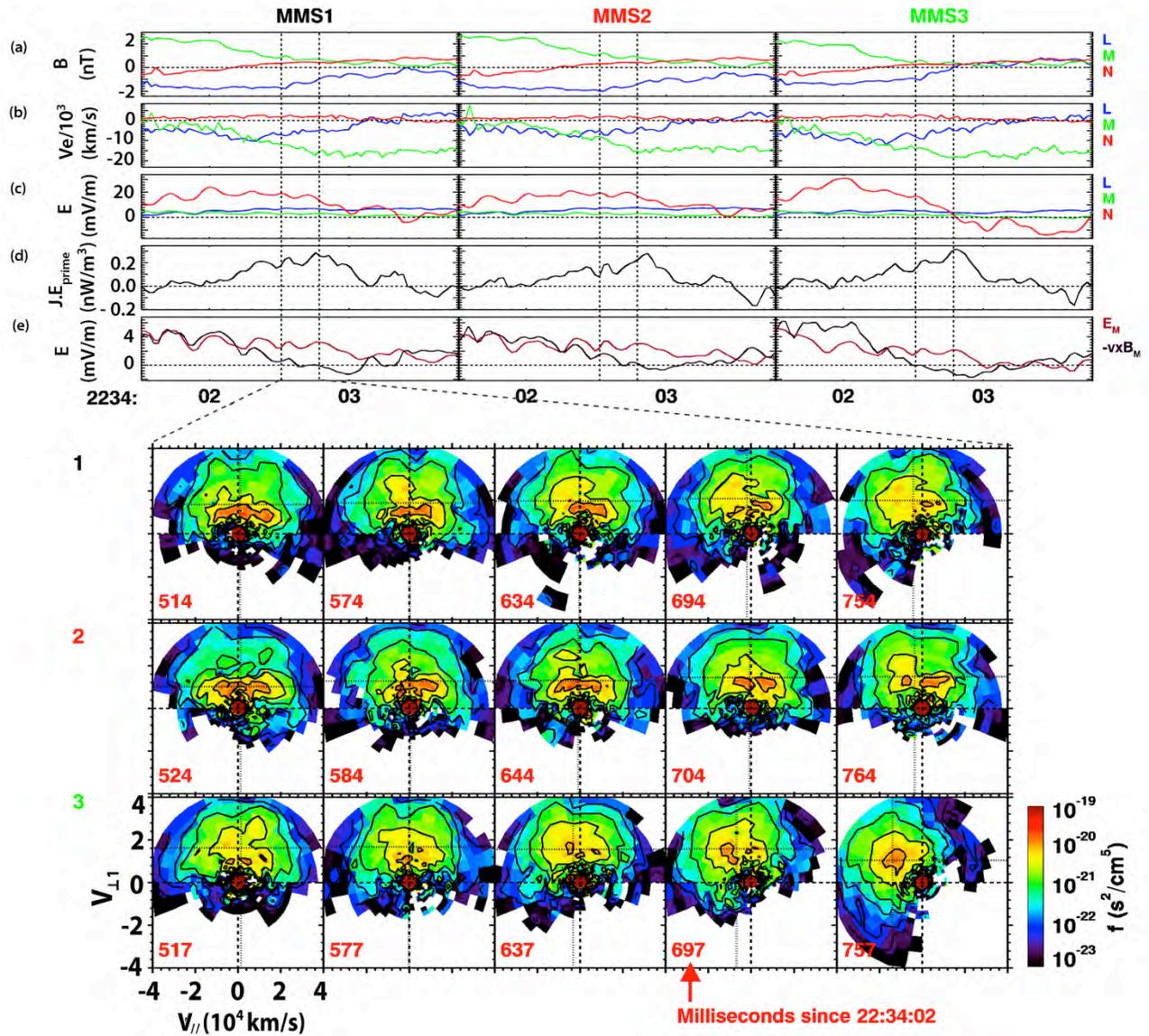

**Figure 3.** Field data and electron DFs for three MMS spacecraft on 2017 July 11, for ~2 seconds around the EDR. (MMS4 resembles MMS2) Upper panels, for each spacecraft:(a) components of **B**; (b) Electron bulk velocity; (c) **E**, where the reversal in $E_N$ is clearly seen on MMS3 and briefly MMS1, but not the others ;  (d) **J · E'** ; (e) M-component of **E** and -( **V**$_e$ x **B**); Lower panels, from 2.604s to 2.784s are the reduced (summed over $V_{\perp 2}$) electron 30 ms DFs in ($V_{||}$, $V_{\perp 1}$) for each spacecraft at the times between dotted lines in upper panels.

**Supplementary Materials**

Supplementary text

Fig. S1 - S3

Movie S1

**SUPPLEMENTARY MATERIALS**

**S1.** Location and Configuration of MMS at 22:34:00 UT on 11 July 2017, and the LMN coordinate system.

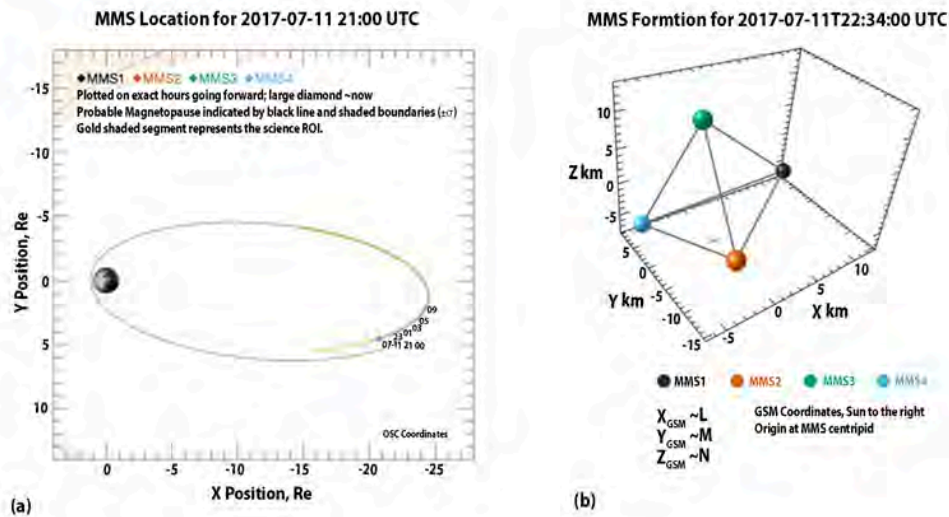

**Figure S1(a)** shows an ecliptic-plane projection of MMS orbit in geocentric-solar-ecliptic (GSE, nearly inertial) coordinates on 2017 July 11. The beige area is the MMS region of interest where high resolution data are taken. Between 21:00 and 22:00 on this date, mild magnetospheric substorm activity began in the magnetotail. Panel (b) shows the MMS spacecraft tetrahedral formation in GSM (Geocentric Solar Magnetospheric) coordinates: **X** towards Sun; **Y** perpendicular to Earth's magnetic dipole and X, pointing towards dusk; **Z = X** x **Y.** A LMN coordinate system for the EDR encounter was established by first determining **N** in the direction of the maximum directional derivative in B at 22:34:02s, then determining the **M** vector perpendicular to this **N**, which maximizes the magnitude of the reconnecting current, as seen in Figure 2e, and **L** then is given by **N**x**M**. [L;M;N] in GSE coordinates are given as [0.971, 0.216, -0.106; -0.234, 0.948, -0.215; 0.054, 0.233, 0.971]. Thus, $X_{GSM} \sim L$ ; $Y_{GSM} \sim M$; and $Z_{GSM} \sim N$, as indicated in the figure. It should be noted that a LMN system can be determined in many different ways, and the several that have been applied for this case are all somewhat similar to the above, but vary by 5-10 degrees. This has a major impact on only one important quantity, $E_M$. Figure 3c shows a normal electric field, $E_N$, ~30 mV/m, whereas $E_M$ (in figures 2g, 3c, and 3e) is ~1-2 mV/m. Thus a change in the N-M axes by only 4 degrees can change $E_M$ by 100%. Due to this fact and the underlying uncertainty of the electric field measurement of ~1 mV/m, an estimate of the reconnection rate from the value of $E_M$ is not reliable. This emphasizes the importance of relying on the scalar quantity, **J** ⋅ **E′** (figures 2h and 3d), which shows clearly that electromagnetic energy is being converted to plasma energy, independent of a particular LMN system.

## S2: Computation Details and Calibration Corrections

The ion velocities in figure 1f have been corrected for the fact that the FPI spectrometer does not cover a sufficient energy range to account for high energy ion phase space when the ion temperature is greater than about 10 keV, as seen in figure 1a. Comparisons over some 5 minutes around 22:34:00 with the other ion spectrometer, HPCA (Hot Plasma Composition Analyzer), which goes to higher energies, and with the **E**x**B**/$B^2$ velocity show that, during the interval of figure 1, a correction by multiplying the MMS SDC data repository values by a factor of 2, gives values of **V**$_{ion}$ that are good to about 20%. In addition, the low density of this EDR encounter resulted in spacecraft potential values greater than 50v. The local photoelectrons from the spacecraft can therefore be seen in figure 1b, at energies below this value. These electrons were removed in the data of figure 2b by plotting energies only greater than 50 eV, and in the computation of the DFs in figures 2 and 3. This same high spacecraft potential affects  offsets of the axial electric field sensor, which then drifts over time scales ~ 10 seconds, the same time scale as the electron density in figure 1e. With running medians over 8 seconds, the offsets of the median electric field were recalibrated to agree with the median of -$\mathbf{V}_e$ x$\mathbf{B}$. The electric field was low pass filter with 3 dB point of 8 Hz to correspond to the approximate time dependence of the electron moments.

Using plasma moments, the time profiles of the current peak, from 22:34:01.8 to ~22:34:04.0 in Figure 2e are nearly identical in all four spacecraft, and agree with the curlometer calculation of the current ($\mathbf{J}_{curl}$) at the barocenter of the MMS tetrahedron, except that $\mathbf{J}_{curl}$ is a constant fraction (0.8) of the plasma $\mathbf{J}$. Since all the profiles are the same in time, the current width must have been greater than the spacecraft separation (~15km). The factor of largest uncertainty is the electron density, which therefore is adjusted to be 0.8 of the SDC value in the calculations below.

A more detailed estimate of the normal half-width follows from Ampere's Law:

$$\mu_0 J_M = (curl(B))_M = (\partial_N B_L - \partial_L B_N ).$$

In our case, $B_N$ is relatively small, and $\partial_L B_N$ even smaller, so we ignore this term. We can thus compute the N position as an integral,

$$N = \int dB_L / \mu_0 J_M$$

over ~1 second from both the rise of the cross-tail current and the appearance of accelerated electrons (~22:34:02.0, Figures 2e,b) to the midplane crossing (i.e. neutral sheet and also current maximum) at 22:34:03.0. This gives a value of 25 km, a little less that the simple division in the main text. A less reliable estimate follows from timing.. As mentioned, the timing analysis at 22:34:02.2s shows the structure with a normal velocity $V_{XN}$ ~ -70 km/s, but falling to zero by 22:34;03.0. An average of these two normal velocities (35 km/s), multiplied by the time elapsed above, ~1s , yields an estimate of the EDR normal half-width of ~35 km .We therefore have bounds on this normal width of certainly greater than 15 km (from current profiles) to a higher estimate of 35km, from timing, with the two from Ampere's Law, 25-30km in the middle. We have chosen the simple Ampere value of 30 km for the half-width.

## S3. Simulation parameters

The 2D Particle-In-Cell (PIC) simulation of figures 2o-r was computed over 3200x3200 grid, a mass ratio $m_i/m_e$=100, and $7x10^{10}$ particles of the type described in (*27*). The initial configuration is a Harris current sheet, with a customary superposed X-type perturbation and a background density of 0.2 in both inflow regions. The ion-electron temperature ratio is 5, and the initial temperature is constant throughout the modeled system. The actual locations of the DFs from this simulation in Figure 2 are: (L=-0.5, N=0.1 for (p,q) and L= 0.5 ,N=0 for (r,s) ).

## S4. Reconnection Topology

The topology of the field lines around the EDR is confirmed by examining pitch angle distributions of the (very) energetic electrons (40-130 keV, Figure **S2**). These data are 40 seconds around the EDR, which is clearly prominent in the electron velocity peak in panel (c).  At 22:34:07UT, just earthward of the EDR, MMS3 saw a burst of energetic electrons in the anti-parallel direction, presumably accelerated in the reconnection process and streaming out along the separatrices. This would be consistent with the return of MMS3 below the neutral sheet on the earthward side of the EDR. Later and further into the exhaust, these electrons are seen filling in the pitch angles (although with enhanced fluxes in the anti-parallel direction) and presumably now trapped in the Earthward extent of the magnetic field. Tailward of the EDR, before 22:34:00, these electrons are not visible, indicating that they had escaped along field lines connected at both ends to the solar wind. This figure is strong evidence not only of magnetic reconnection on-going at the EDR, but also of the ability of the reconnection process to accelerate electrons to very high energies, one of the long standing features associated with reconnection in solar flares and other astrophysical situations.

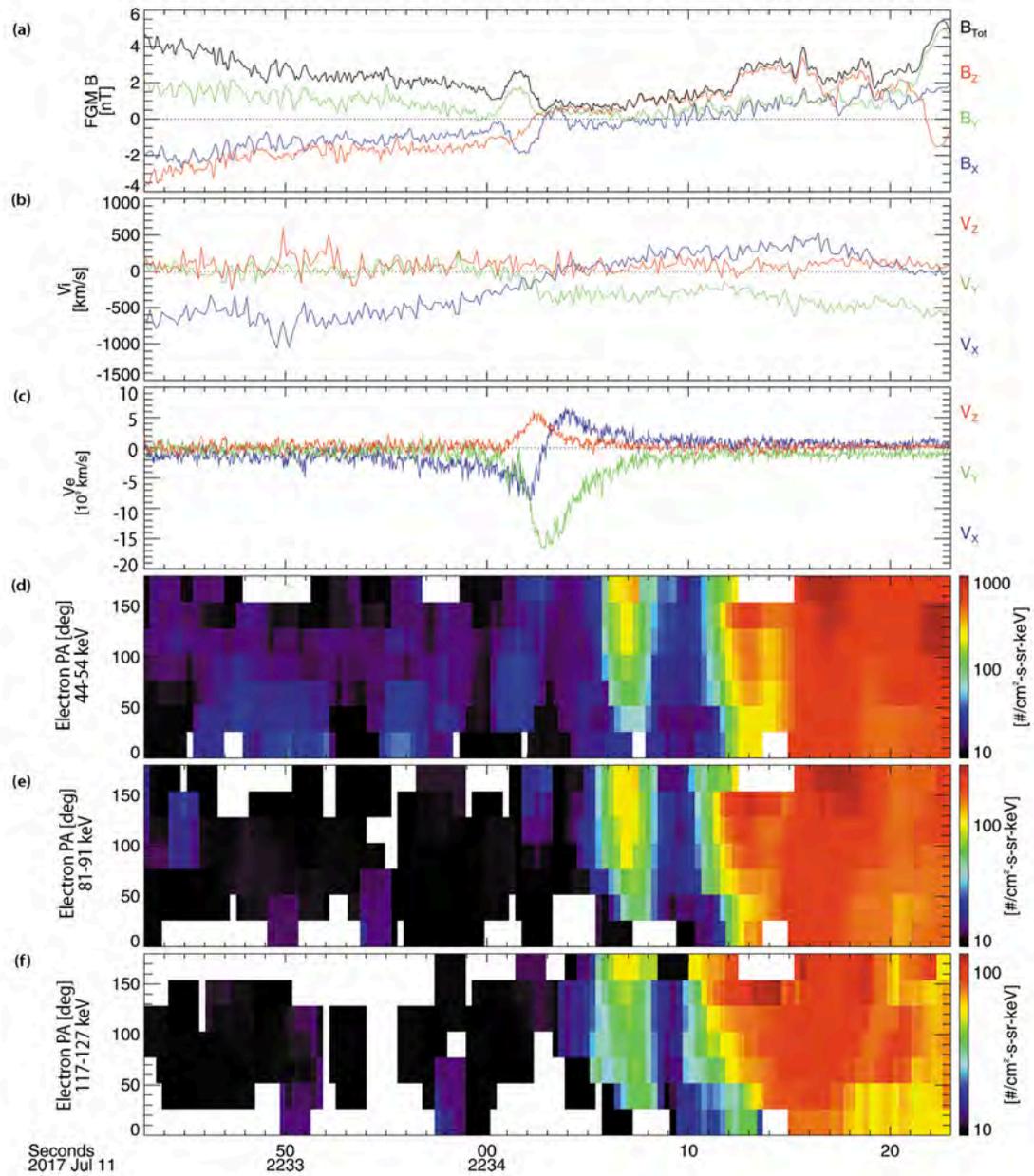

**Figure S2**. MMS3 magnetic field(a), Ion (b) and electron (c) velocities, and electron pitch-angle distributions (d-f) for three selected high energies.

**Movie S1: Electron Velocity-space Distributions**

**Movie S1** shows a 6-second segment of burst-mode electron distributions keyed to a plot of plasma and field data covering the same time period as Figure 2, but from all MMS spacecraft. These distributions are accumulated over four 30ms electron DFs to bring out detailed features when the density is as low as $0.03 \text{cm}^{-3}$, as in this case. The velocity axes are in the same LMN coordinate system as described in S1. Even with a four-fold accumulation, the increase in electron time resolution is an important reason why MMS is able to investigate the electron-scale physics of reconnection for the first time.

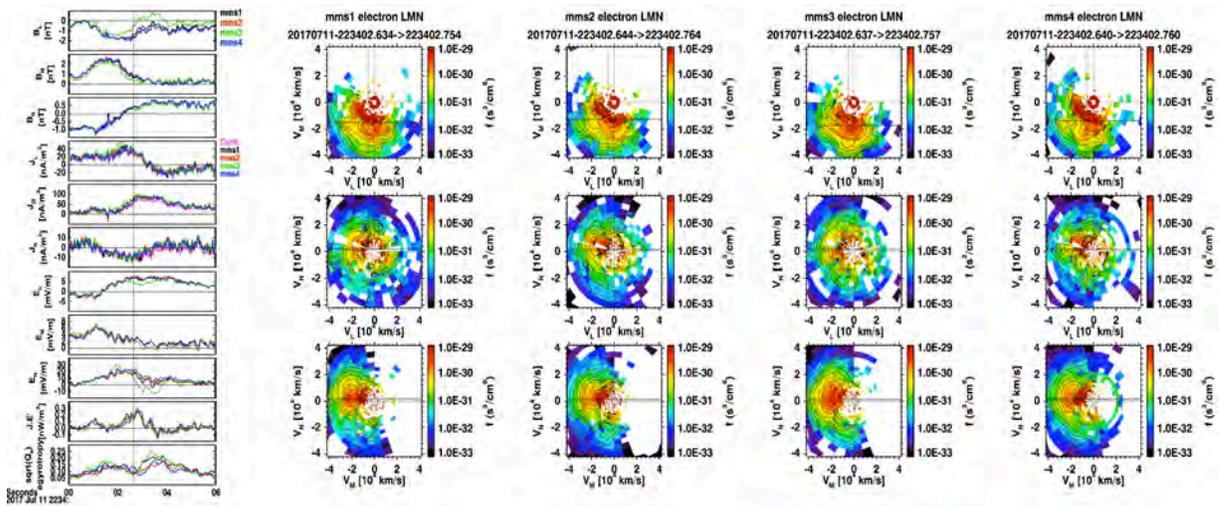

**Figure S3:** One frame of movie **S1**, showing in the left panel, the time history (from top to bottom) of : $B_L$, $B_M$, $B_N$, $J_L$, $J_M$, $J_N$, $E_L$, $E_M$, $E_N$, **J** $\cdot$ **E′**, and the Swisdak agyrotropy index for all four

spacecraft. The right four columns show the reduced electron distributions for each spacecraft along the indicated LMN velocities at the time of the vertical dash line in the left panel.

**S5**: Additional Acknowledgements:

Contributions from (1) Imperial College London, supported by the STFC(UK) grant ST/N000692/1; (2) from Institut de Recherche en Astrophysique et Planétologie, supported by CNES, CNRS-INSIS and CNRS-INSU.